\documentstyle[12pt]{article}
\begin{document}
\begin{center}
{\large \bf Summary of the
Theory Part}\\
{\large \bf of the International Conference on Elastic and Diffractive
Scattering - }\\
{\large \bf "Frontiers in Strong Interactions" (VIth Blois Workshop),}\\
{\large \bf Chateau de Blois, France, June 20 -24, 1995}\\
\bigskip
J. Kwieci\'nski\\
Department of Theoretical Physics\\
H. Niewodnicza\'nski Institute of Nuclear Physics\\
ul. Radzikowskiego 152, 31 342 Krak\'ow, Poland.
\end{center}
\bigskip
\begin{abstract}
The theory part of the conference is summarized with certain emphasis on the
results concerning the pomeron in soft and hard processes.
\end{abstract}
When summarizing the content of the meeting which embraced the very wide
spectrum of subjects ranging from the classical dispersion relation
analysis of forward scattering to the theory of hard processes
based on perturbative QCD it is useful to find some "common
denominator" of as many presentations as possible.  I believe that
a possible choice in this case might be the pomeron structure in soft
and hard processes.\\

Substantial part of this conference has been as usual devoted
 to the
analysis of the forward scattering amplitude parameters (i.e. of
$\sigma_{tot}(s)$ and $\rho$) in a model independent way assuming only
dispersion relations and various forms of the asymptotic parametrisations of
the cross-sections which respected the Froissart-Martin bound
and other asymptotic theorems.
The theoretical discussion of the forward scattering amplitude has been
presented in
\cite{BLOCK,GAURON,KKANG,SKKIM,AMARTIN,KHURI,KUNDRAT,SELYUGIN}.  The data
seem to support
the dominance of the crossing-even amplitude and there is little evidence
for a significant odderon contribution in forward and near forward
scattering.  The real part just reflects (through the dispersion relations)
the increase of the total cross-sections with increasing energy.\\

 We have had numerous contributions to this conference
discussing various aspects of the pomeron physics ranging from the pure
phenomenology to the discussion of formal problems of  the pomeron
singularity in perturbative QCD.\\

The term "pomeron" corresponds to the mechanism of diffractive scattering at
high energy. It is relevant for the description of several
phenomena and quantities like the total cross-sections $\sigma_{tot}(s)$
and their energy
dependence, the real part of the scattering ampltude,
the variation with energy of the differential elastic cross-section
$d\sigma/dt$,
behaviour of the diffractive cross-section $d\sigma/dtdM^2$, behaviour of
the deep-inelastic scattering structure function $F_2(x,Q^2)$ at low $x$,
behaviour of the diffractive structure function etc.  \\

The simplest yet presumably very incomplete description of the pomeron
is within the Regge pole model.
In this model one assumes  that a pomeron is described by a Regge pole
with the  trajectory $\alpha_P(t) =
\alpha_P(0)+\alpha^{\prime}t$. The scattering amplitude
corresponding to the pomeron
exchange is  given by the following formula:
\begin{equation}
A(s,t)= -g(t) {exp(-i{\pi \over 2}\alpha_P(t))\over
sin({\pi \over 2}\alpha_P(t))}s^{\alpha_P(t)}
\label{regge}
\end{equation}
where the function $g(t)$ describes the pomeron coupling.
Using the optical theorem one gets the following high energy
behaviour of the total
cross-sections:
\begin{equation}
\sigma_{tot}(s) \sim {Im A(s,0)\over s} = g(0)s^{\alpha_P(0)-1}
\label{sigtot}
\end{equation}
One also gets:
\begin{equation}
\rho ={ReA(s,0)\over ImA(s,0)}=ctg({\pi \over 2}\alpha_P(0))
\label{rho}
\end{equation}
with corrections from low lying Regge trajectories which vanish at high
energies approximately as $s^{-1/2}$.
It follows from (\ref{sigtot}) that the Regge-pole model of a pomeron can
describe the increase of the total
cross-sections with energy assuming $\alpha_P(0)
> 1$ but this parametrization will eventually violate the Froissart-Martin
bound.\\

Phenomenological description of $\sigma_{tot}$ and of $d\sigma/dt$ proceeds
in general along the following two lines:
\begin{enumerate}
\item
One introduces the "effective" (soft) pomeron with relatively low value
of its intercept ($\alpha_P(0) \approx 1.08$ \cite{DONNACHIE}) which
can very well describe the high energy behaviour of all hadronic and
photoproduction cross-sections (with possible exception of the one
CDF point).  In phenomenological analysis one also adds the reggeon
contribution which gives the term $\sigma_{tot}^R \sim s^{\alpha_R(0)-1}$
with $\alpha_R(0) \approx 0.5$.\\
The power-like increase of the total cross-section has to be, of course,
slowed down
at asymptotic energies but  those corrections
are presumably still relatively unimportant at presently available energies.
It has however
been observed \cite{MAOR} that the soft pomeron can violate unitarity
for central $pp$ partial wave already for $\sqrt{s}>2.5TeV$.
\item
One considers from the very beginning the unitarized amplitude using
the eikonal model \cite{MAOR,TTWU,DAKHNO}.  In this model
the partial wave amplitude   $f(s,b)$ has the form
\begin{equation}
f(s,b)={1\over 2i}[exp(-2\Omega(b,s))-1]
\end{equation}
\begin{equation}
\Omega(s,b)=h(b,s)s^{\Delta}{exp(-i{\pi \over 2}\Delta)\over
cos({\pi \over 2}\Delta)}
\label{eikonal}
\end{equation}
where $\Delta >0$ and $h(b,s)$ is the slowly varying function of $s$.
The variable $b$ denotes the impact parameter. The eikonal function can be
viewed upon as originating from the "bare" pomeron with its intercept
$\alpha_P(0)=1+\Delta$ being above unity. The eikonal models
with  $\Omega(s,b) \sim s^{\Delta} (\Delta > 0)$ lead to the scattering on
the expanding (black) disk at asymptotic energies. The radius of the disk
grows logarithmically with increasing energy (i.e. $R(s) \sim ln(s)$ ). This
leads to the saturation of the Froissart-Martin bound at asymptotic energies
and to the  deviation of the shape of the
diffractive peak from the simple exponential.  Inelastic diffractive
scattering becomes peripheral at asymptotic energies \cite{MAOR}.
The recent fits presented in this conference $\cite{MAOR,DAKHNO}$ give
rather large values of $\Delta$ ($\Delta \approx 0.3$ or so).  This parameter
had a relatively small value $\Delta \approx 0.08$) in the  original
eikonal model of Bourely, Soffer and Wu formulated more then 20 years ago
(see \cite{TTWU} and references therein). Other aspects of the
eikonal model (called also the Chou-Yang model) have been presented
in \cite{ISLAM,FAZEL}.
\end{enumerate}

The Regge phenomenology may also be applicable in the analysis of the deep
inelastic scattering in the limit when the Bjorken variable $x$ is small.
The inelastic lepton scattering (i.e. the reaction $l(p_l)+p(p) \rightarrow
l^{\prime}(p_l^{\prime}) + anything$) is related through the one photon
exchange
approximation to the forward virtual Compton scattering $\gamma^*(Q^2)+
p(p) \rightarrow \gamma^*(Q^2)+p(p)$ where $Q^2=-q^2$, $q=p_l-p_l^{\prime}$
and $x=Q^2/2pq$.  The measured structure functions $F_2(x,Q^2)$ and
$F_L(x,Q^2)$ are directly related to the total (virtual) photoproduction
cross-sections $\sigma_T$ and $\sigma_L$ corresponding to transversely
and longitudinally polarized photons:
\begin{equation}
F_2(x,Q^2)={Q^2\over 4 \pi^2 \alpha}(\sigma_T +\sigma_L)
\label{fsigma2}
\end{equation}
\begin{equation}
F_L(x,Q^2)={Q^2\over 4 \pi^2 \alpha} \sigma_L
\label{fsigmal}
\end{equation}
Assuming the conventional Regge pole parametrisation for $\sigma_{T,L}$
\begin{equation}
\sigma_{T,L}={4\pi^2 \alpha \over Q^2}\sum ({2pq\over Q^2})^{\alpha_i(0)-1}
C_{T,L}^i(Q^2)
\label{sregge}
\end{equation}
one gets  the following small $x$ behaviour for the structure functions:
\begin{equation}
F_{2,L}=\sum (x)^{1-\alpha_i(0)}C_{T,L}^i(Q^2)
\label{fregge}
\end{equation}
where the sum in (\ref{sregge},\ref{fregge}) extends over the pomeron and
the reggeon contributions.  The experimental results from HERA show  that
the structure function $F_{2}(x,Q^2)$ for moderate and large $Q^2$ values
($Q^2 >$ 1.5 $GeV^2$ or so) grows more rapidly then expected on the basis of
the
straightforward extension of the Regge pole parametrization with the
relatively small intercept of the effective pomeron ($\alpha_P(0)
\approx$ 1.08) {\cite{BARTEL}.\\

Very steep behaviour of parton distributions and of structure function
$F_2(x,Q^2)$ has been predicted long time ago within the perturbative
QCD and  several talks in this conference were devoted to the discussion
of the small $x$ physics in perturbative QCD
\cite{DONNACHIE,KOLYA,GUSTAFSON,CIAFALONI,CITAN,LIPATOV,BRAUN,FADIN,WHITE}
(see also \cite{PETROV,MARTYNOV,FORTE1,KOTIKOV}).\\

The relevant framework for discussing the small $x$ limit of parton
distributions is the leading log$1/x$ (LL$1/x$) approximation which
corresponds to the sum of those terms in the perturbative expansion
where the powers of $\alpha_s$ are accompanied by the leading
powers of ln($1/x$).  At small $x$ the dominant role is played by the
gluons and the quark (antiquark) distributions as well as the deep
inelastic structure functions $F_{2,L}(x,Q^2)$ are also driven by the
gluons through the $g \rightarrow q\bar q$ transitions.  \\

The basic quantity at small $x$ is the unintegrated gluon distribution
$h(x,k^2)$ which satisfies the following equation:
\begin{equation}
-x{\partial h(x,k^2)\over \partial x}=
\bar \alpha_s \int {d^2q\over \pi q^2}[h(x,(\hat q + \hat k)^2)-
h(x,q^2)\Theta(k^2-q^2)]
\label{bfkl}
\end{equation}
where $\bar \alpha_s=3\alpha_s/\pi$.
This equation is the celebrated Balitzkij, Fadin, Kuraev, Lipatov (BFKL)
equation (see \cite{KOLYA} and references therein) which
corresponds to the
sum of ladder diagrams with gluon emissions.  The second term at the right
hand side of the eq.(\ref{bfkl}) describes the virtual corrections. The
longitudinal momenta are strongly ordered along the chain yet the
transverse momenta are not ordered.
The two-dimensional vector $\hat k$ describes the transverse momentum of the
gluon while the vector $\hat q$ is the transverse momentum of the
produced gluon "jet" at the last rung of the chain.
The familiar (scale dependent) gluon distribution $g(x,Q^2)$ is
related in the following way to the unintegrated distribution $h(x,k^2)$
\begin{equation}
xg(x,Q^2)=\int^{Q^2} dk^2h(x,k^2)
\label{intg}
\end{equation}
After resumming the
"unresolvable" real emission ($q^2 < \mu^2$) and virtual corrections
one can rearrange the BFKL equation into the following form:
\begin{equation}
h(x,k^2)=h^0(x,k^2)+ \bar \alpha_s\int_x^1{dz\over z}z^{\omega(k^2,\mu^2)}
\int {d^2q\over \pi q^2}h({x\over z},(\hat q + \hat k)^2)
\label{bfklf}
\end{equation}
where
\begin{equation}
\omega(k^2,\mu^2)=\bar \alpha_s ln({k^2\over \mu^2})
\label{omega}
\end{equation}
The equation (\ref{bfklf}) sums now only the real resolvable radiation.
The quantity $\omega(k^2,\mu^2)$ is directly related to the gluon
Regge trajectory $\alpha_g(k^2,\mu^2)$
\begin{equation}
\omega(k^2,\mu^2)=2(1-\alpha_g(k^2,\mu^2))
\label{glut}
\end{equation}
The equation (\ref{bfklf}) corresponds to the ladder diagrams with the
reggeized gluon exchange along the ladder.  One can also interpret
the damping factor $z^{\omega(k^2,\mu^2)}$ which correspond to the
reggeized gluon exchange as the "non-Sudakov form-factor" $\Delta_{NS}$
\begin{equation}
\Delta_{NS}=z^{\omega(k^2,\mu^2)}=exp\left(-\bar \alpha_s \int_z^1{dz^{\prime}
\over z^{\prime}} \int_{\mu^2}^{k^2} {dk^{\prime 2}\over k^{\prime 2}}\right)
\label{nsl}
\end{equation}

The asymptotic solution of the BFKL equation in the small $x$ limit has the
following form:
\begin{equation}
k^2h(x,k^2) = C {({k^2\over k_0^2})^{1/2}\over ln^{1/2}(1/x)}
x^{-\lambda}exp\left(-{ln^2({k^2\over k_0^2})\over
2\lambda^{\prime \prime}ln(1/x)}\right)
\label{bfkls}
\end{equation}
where
\begin{equation}
\lambda=\bar \alpha_s 4 ln2
\label{bfkli}
\end{equation}
and $\lambda^{\prime \prime} = 28 \bar \alpha_s \zeta(3)$ with the
Riemann zeta function $\zeta(3) \approx$ 1.202.  The solution of the BFKL
equation exhibits very strong increase with decreasing $x$ when compared
with the increase implied by the "soft" Pomeron since $\lambda >> 0.08$
for reasonable choice of the coupling $\bar \alpha_s$. The pomeron
which is connected with the solution of the BFKL equation is usually referred
to as the "hard" pomeron.  It is expected to control the small $x$ behaviour
of (semi)hard processes.\\

The small $x$ increase is correlated with the increase in $k^2$ and with the
diffusion pattern of the solution of the BFKL equation.  This is closely
related to absence of transverse momentum ordering along the gluon chain
and has several implications for the structure of the final state in deep
inelastic scattering.  There are several measurements which are aimed
at revealing this mutual relationship between small $x$ increase and the
increase of transverse momentum.  They are (for instance) measurements
of energetic jets in deep inelastic scattering, measurements of the energy
flow in the central region, measurements of azimuthal decorrelation of
dijets etc. Several new experimental results
concerning the study of the deep inelastic final state from the point of view
of revealing the BFKL signals have been reported
in this conference \cite{BARTEL,HARTMANN}.\\

In the impact parameter representation the BFKL equation offers an
interesting
interpretation in terms of the colour dipoles and this approach
has been summarized in the talk by Kolya Nikolaev \cite{KOLYA}.
Possible generalization of the BFKL formalism which includes energy
momentum conservation constraint and the coherence effects was
discussed by Gosta Gustafson \cite{GUSTAFSON}.  \\

Observable quantities at small $x$ are calculated in terms of the solution
of the BFKL equation using the $k_t$ factorization theorem.  The deep
inelastic lepton hadron scattering is dominated at small $x$ by the
gluon-photon fusion and the $k_t$ factorization formula
for the structure  functions $F_{2,L}(x,Q^2)$
then reads::
\begin{equation}
F_{2,L}(x,Q^2)=\int_{x}^1{dz\over z} \int dk^2 \hat F^0_{2,L}({x\over z},k^2,
Q^2)h(k^2,z)
\label{ktfac}
\end{equation}
In this equation the functions $\hat F^0_{2,L}({x\over z},k^2,
Q^2)$  are the structure functions of the off-shell gluon of virtuality $k^2$
and correspond to the quark box contribution to the gluon-photon fusion
process.  The unintegrated gluon distribution $h(k^2,z)$ is the solution
of the BFKL equation.  The small $x$ behaviour of $F_{2,L}(x,Q^2)$
reflects the small $z$ behaviour of $h(k^2,z)$ i.e. the structure functions
at small $x$ are driven by the gluon.\\

The leading twist part of the $k_t$ factorisation formula can be rewritten in
a collinear factorization form.  The leading small $x$ effects are then
automatically resummed in the corresponding anomalous dimensions (or
splitting functions) and in the coefficient functions. They can
also affect the (nonperturbative) starting distributions.
In this way one can systematically include the leading ln($1/x$) effects
within conventional formalism based on the Altarelli-Parisi evolution
equations for parton densities and collinear factorization formulas
for calculating the observable quantities.  All these (and related)
problems have been nicely summarized in the review talk given by Marcello
Ciafaloni \cite{CIAFALONI}.\\

The possible role of the leading log$1/x$ resummation which go beyond the
standard leading (or next-to-leading log($Q^2$)) QCD analysis is still
not very well understood. In particular one can get equally good description of
HERA data on $F_2(x,Q^2)$ using the QCD evolution formalism
without those small $x$ "corrections",  provided that
the starting parton distributions are changed appropriately.  The
structure function $F_2(x,Q^2)$ is therefore not the best
"discriminator" of the BFKL small $x$ effects and it is the dedicated
study of final states in deep inelastic scattering (along the lines
described above) which may prove to be a very useful tool for this
purpose.  The HERA data have nevertheless put important constraints
on the small $x$ behaviour of deep inelastic scattering which should
have implications also in the kinematical range beyond that which
is currently accessible.  Possible implications of the QCD expectations
for the small $x$ behaviour of deep inelastic scattering for the
estimate of ultra -high energy neutrino cross sections has
been presented by Ina Sarcevic \cite{SARCEVIC}. \\

Several new interesting results have been reported which concern the formal
studies of the high energy (or small $x$) limit in QCD beyond the
leading logarithmic approximation \cite{LIPATOV,BRAUN,FADIN,WHITE}.
Important theoretical tool
in this case is the effective field theory where the basic objects are
the   reggeized gluons and the effective action of this effective field
theory
obeys conformal invariance \cite{LIPATOV}. Theoretical analysis
simplifies in the large $N_c$ limit and the theory resembles then the
Ising model. One can discuss both the pomeron
which appears as the bound state of two (reggeized) gluons, the
odderon (i.e. the bound state of three reggeized gluons) as well as bound
states of many reggeized gluons.\\

The next-to-leading ln$(1/x)$ corrections can be present in all relevant
quantities i.e. in the particle-particle-reggeon vertex,
the reggeon-reggeon-particle vertex and in the gluon Regge trajectory.
(The reggeon here corresponds to the reggeized gluon).  Besides that
one has also to include  additional region of phase-space
which goes beyond strong ordering of longitudinal momenta. The next-to-leading
corrections to the gluon trajectory have been presented at this conference
by Victor Fadin \cite{FADIN} and the results concerning those corrections
to the BFKL kernel were reported by Alan White \cite{WHITE}.\\

A large portion of the conference has been devoted to discussion of the deep
inelastic diffraction
\cite{DONNACHIE,KOLYA,KAIDALOV,GENOVESE,DIEHL,ABRAMOWICZ,DAINTON} with
the basic theoretical issues of diffraction  summarized in the
review talk given by Alosha Kaidalov \cite{KAIDALOV}. It should also
be reminded that the ideas concerning possible partonic content of the
pomeron were for the first time reported by Ingelman and Schlein exactly
ten years ago at the first Blois Conference in 1985.\\

The deep inelastic diffraction is a process:
\begin{equation}
l(p_l)+p(p) \rightarrow l^{\prime}(p_l^{\prime}) + X +p^{\prime}(p^{\prime})
\label{disdif}
\end{equation}
where there is a large rapidity gap between the recoil proton
(or excited proton) and the hadronic system $X$.  To be precise the
process (\ref{disdif}) reflects the diffractive disssociation
of the virtual photon.  Diffractive dissociation is described by the following
kinematical variables:
\begin{equation}
\beta={Q^2\over 2 (p-p^{\prime})q}
\end{equation}
\begin{equation}
x_P={x\over \beta}
\end{equation}
 \begin{equation}
t= (p-p^{\prime})^2.
\label{difv}
\end{equation}
Assuming that the diffraction dissociation is dominated by the pomeron
exchange and that the pomeron is described by
a Regge pole one gets the following factorizable expression for the
diffractive structure function:
\begin{equation}
{\partial F_2^{diff}\over \partial x_P \partial t}= f(x_P,t)F_2^P(\beta,Q^2,t)
\label{difsf}
\end{equation}
where the "flux factor" $f(x_P,t)$ is given by the following formula
:
\begin{equation}
f(x_P,t)=N{B^2(t)\over 16\pi} x_P^{1-2\alpha_P(t)}
\label{flux}
\end{equation}
with $B(t)$ describing the pomeron coupling to a proton and $N$ being the
normalisation factor.  The function $F_2^P(\beta,Q^2,t)$
is the pomeron structure function which in the (QCD improved) parton model
is related in a standard way to the quark and antiquark distribution
functions in a pomeron.
\begin{equation}
F_2^P(\beta,Q^2,t)=\beta \sum e_i^2[q_i^P(\beta,Q^2,t)+ \bar
q_i^P(\beta,Q^2,t)]
\label{f2pom}
\end{equation}
with $q_i^P(\beta,Q^2,t)=\bar q_i^P(\beta,Q^2,t)$.  The variable
$\beta$ which is the Bjorken scaling variable appropriate for
deep inelastic lepton-pomeron "scattering" has now a meaning of the
momentum fraction of the pomeron carried by the
probed quark (antiquark).  The parton distributtions in a pomeron
are assummed to obey the standard Altarelli-Parisi evolution equations:
\begin{equation}
Q^2{\partial q^P\over \partial Q^2}=P_{qq} \otimes q^P + P_{qg} \otimes g^P
\label{app}
\end{equation}
with similar equation for the gluon distribution in a Pomeron.  The first
term in the right hand side of the eq. (\ref{app}) becomes negative
at large $\beta$ while the second term stays always positive and
is ussually very small at large $\beta$ unless the gluon distributions
are large and have a hard spectrum.  The data suggest that the slope
of $F_2^P$ as the function of $Q^2$ does not change sign even
at relatively large values of $\beta$ that favours the hard
gluon spectrum in a pomeron \cite{KAIDALOV,ABRAMOWICZ,DAINTON}.
This should be contrasted with the behaviour of the structure function
of the proton which , at large $x$, decreases with increasing $Q^2$.
The data on inclusive diffractive production do also favour the soft pomeron
with relatively low intercept.  The diffractive production of vector mesons
does seem to require "hard" pomeron contribution
\cite{DONNACHIE,WHITMORE,KENYON}. It has also
be pointed out  that he factorization property
(\ref{difsf}) may
not hold in models based
entirely on perturbative QCD when the pomeron is represented
by the BFKL ladder \cite{KOLYA,GENOVESE,BERERA}. \\

The phenomenological as well as purely theoretical study summarized above
indicate that we may have two pomeron contributions: the "soft" pomeron
which is relevant for describing the soft processes and the "hard" one
which shows-up at the processes characterized by a hard scale.
This simple additive picture has however been questioned in some  of
the contributions to this conference \cite{PREDAZZI}.\\

Several contributions have been devoted to the spin effects
in high energy scattering
\cite{BUTTIMORE,FORTE2,BOURRELY,BERTINI,CONTOGOURIS}
and in particular to the disscussion
of the polarized structure functions.
The present theoretical situation in this field has been
summarized  by Stefano Forte
\cite{FORTE2}.   \\

The first moment of the (polarized) structure function $g_1(x,Q^2)$ is
directly related to the matrix elementa of the axial current(s).  For the
isotriplet structure function one gets the Bjorken sum rule:
\begin{equation}
\int_0^1dx(g_1^p(x,Q^2)-g_1^n(x,Q^2)) ={1\over 6}g_A C^{NS}(Q^2)
\label{bjsr}
\end{equation}
where
\begin{equation}
C^{NS}(Q^2)=1-{\alpha_s(Q^2)\over \pi}-3.58({\alpha_s(Q^2)\over \pi})^2-
20.22({\alpha_s(Q^2)\over \pi})^3+.....
\label{cns}
\end{equation}
Analysis of the Bjorken sum rule can give an (independent) determination
of $\alpha_s$ and the result of this analysis is:
\begin{equation}
\Lambda_{n_f=3}^{\bar M \bar S}=383^{+126}_{-116} MeV
\label{lambda}
\end{equation}
\begin{equation}
\alpha_s(M_Z^2)=0.118 \pm 0.007
\label{alphamz}
\end{equation}
The sum rules for the singlet structure function are sensitive through
the axial anomaly on the polarized gluon distribution $\Delta G$
in a proton. \\

At small $x$ the QCD corrections can be very important and can change the
 Regge pole like behaviour.  The canonical Regge pole
expectations is that the polarized structure function $g_1$ should
have the power-like behaviour:
\begin{equation}
g_1(x,Q^2) \sim x^{-\alpha_A(0)}
\label{smxg1}
\end{equation}
where $\alpha_A(0)$ is the intercept of the Regge trajectory corresponding
to axial vector mesons (i.e. $A_1$ Regge trajectory for the non-singlet case
etc.).  It is expected that those Regge trajectories should have relatively
low values of their intercepts (i.e. $\alpha_A(0) \le 0$).  This behaviour
is however unstable against the QCD evolution which generates steeper
$x$ dependence:
\begin{equation}
g_1^{NS}(x,Q^2) \sim exp(c_1 \sqrt{\xi(Q^2)ln(1/x)})
\end{equation}
\begin{equation}
g_1^{S}(x,Q^2) \sim exp(c_2 \sqrt{\xi(Q^2)ln(1/x)})
\label{smxqcd}
\end{equation}
where
\begin{equation}
\xi(Q^2)=\int^{Q^2} {dq^2\over q^2} {\alpha_s(q^2)\over \pi}
\end{equation}
and where  $c_1<c_2$ because of the $gg\rightarrow q \bar q$ mixing in
the QCD evolution
of polarized singlet densities.  Several contributions to the conference
discussed specific models for polarized parton denssities with QCD evolution
included \cite{BOURRELY,BERTINI}.\\

Besides the topics summarized above several contributions discussed
various issues of multiparticle production
in hadronic collissions \cite{RANFT,FIALKOWSKI,FERREIRA}, various
QCD tests including the review on fragmentation functions \cite{GUILLET} etc.
  One should finally
mention an interesting presentation by Elena Papageorgiou on the
the signals of new phase of QED in elastic scattering on nuclei
\cite{PAPAGEORGIOU}. \\

To sum up we have witnessed enormous progress in both phenomenological
as well as the theoretical understanding of the pomeron.
Change of its nature with relevant scales is evidently visible in the
data and needs satisfactory dynamical explanation.
\vskip1cm
Acknowledgments.\\

I would like to thank Maurice Haguenauer for his kind invitation to take part
at this Conference and for very warm hospitality in Blois.
This research has been
supported in part by the Polish State Committe for Scientific Research
grant N0 2 P302 062 04.

\end{document}